\documentclass[a4paper, amsfonts, amssymb, amsmath, reprint, showkeys, nofootinbib, twoside,aps, pra]{revtex4-2}
\usepackage[dvipsnames,table,xcdraw]{xcolor} 
\usepackage[english]{babel}
\usepackage[utf8]{inputenc}
\usepackage[colorinlistoftodos, color=green!40, prependcaption]{todonotes}

\usepackage{amsthm}
\usepackage{mathtools}
\usepackage{physics}
\usepackage{graphicx}
\usepackage[left=23mm,right=13mm,top=35mm,columnsep=15pt]{geometry} 
\usepackage{adjustbox}
\usepackage{placeins}
\usepackage{siunitx} 
\usepackage[T1]{fontenc}
\usepackage{lipsum}
\DeclareGraphicsExtensions{.pdf,.png,.jpg,.eps}
\usepackage{csquotes}
\usetikzlibrary{decorations.markings}
\usepackage{tikz} 
\definecolor{circle_gray}{RGB}{100,100,100}
\definecolor{ellipse_gray}{RGB}{150,150,150}
\definecolor{linear_gray}{RGB}{200,200,200}
\usetikzlibrary{calc,fadings,decorations.pathreplacing}
\usetikzlibrary{decorations.pathreplacing,decorations.pathmorphing}
\usetikzlibrary{arrows}

\usepackage{hyperref} 
\hypersetup{
	colorlinks=true,
	linkcolor=blue,
	citecolor=blue,
	filecolor=blue,
	urlcolor=blue,
}
\usepackage{color}
\usepackage{ulem}

\newcommand\old{\bgroup\markoverwith{\textcolor{ForestGreen}{\rule[0.5ex]{2pt}{0.8pt}}}\ULon}

\begin{document}
	
	\title{Effective two-level  approximation of a multi-level system driven by coherent and incoherent fields}
	
	\author{R. Veyron}
	\author{V. Mancois}
	\author{J.B. Gerent}
	\author{G. Baclet}
	\author{P. Bouyer}
	\author{S. Bernon}
	
	\email[Correspondence email address: ]{simon.bernon@institutoptique.fr}
	\affiliation{ LP2N, Laboratoire Photonique, Num\'{e}rique et Nanosciences, Universit\'{e} Bordeaux-IOGS-CNRS:UMR 5298, rue F.Mitterrand, F-33400 Talence, France}
	
	\begin{abstract}
		
		The numerical simulation of multiple scattering in dense ensembles is the mostly adopted solution to predict their complex optical response. While the scalar and vectorial light mediated interactions are accurately taken into account, the computational complexity still limits current simulations to the low saturation regime and ignores the internal structure of atoms. Here, we propose to go beyond these restrictions, at constant computational cost, by describing a multi-level system (MLS) by an effective two-level system (TLS) that best reproduces the coherent and total scattering properties in any saturation regime. The correspondence of our model is evaluated for different experimentally realistic conditions such as the modification of the driving field polarization, the presence of stray magnetic fields or an incoherent resonant electromagnetic field background. The trust interval of the model is quantified for the D$\phantom{}_2$-line of $^{87}$Rb atoms but it could be generalized to any closed transition of a multi-level quantum system.
		
	\end{abstract}
	\date{\today}
	\pacs{32.80.-t, 32.80.Cy, 32.80.Bx, 32.80.Wr}
	\keywords{multi-level atom, scattering rates, saturation, coherent field, incoherent background}
	\maketitle
	
	\section{\label{sec:intro} Introduction} 
	The response of dense ensembles to coherent optical illumination is a paradigmatic situation to study multiple scattering dynamics in which collective effects can be prominent. They lead, for instance, to modifications of the scattering properties such as line shifts and broadening \cite{Jennewein18} in 1D \cite{Glicenstein20} and 2D systems \cite{Corman2017}, sub- \cite{Guerin16a} and super-radiance \cite{Cottier2018}, the optical phase profile engineering \cite{Jenkins12}  to control the reflection properties \cite{Rui20} of a single atomic layer or the localization of light in different regimes \cite{Cottier19,Wiersma97}.
	Simulations of the coupled dipole equations in the linear-optics regime include  interference effects such as coherent backscattering \cite{Chabe2014,Sokolov2019}{which was also} predicted using  random walk simulations  including the atomic internal structure complexity \cite{Labeyrie03,Labeyrie2000}.  
	
	In contrast with the preceding cases, when increasing the saturation parameter, the system deviates from its linear response \cite{EspiritoSanto2020}, requiring a full quantum treatment that scales dramatically with the atom number. An ensemble of $N$ Multi-Level Systems (MLS) with $k$-levels each yields the diagonalization of a  $k^{N} \times k^{N}$ matrix and is {computationally} out-of-range  when considering more than a few particles.
	
	In a mean-field approach where entanglement between atoms is neglected, the full density matrix can be factorized as the product state of single atom density matrices reducing the matrix dimensions to $kN \times kN $ thus improving the simulation capabilities up to few thousand particles.  For a given computational power, the traditional trade-off for numerical simulations is either to consider the internal atomic structure \cite{Miniatura2002,Lee2016} while reducing the number of particles or to model real atoms by two-level systems (TLS) \cite{Chomaz2012}. The latter allows a quantitative comparison with experiments only in specific situations where the experimental conditions allow one to suppress spurious transitions \cite{Jennewein18}. It represents a loss of generality and a limitation for the general comparison of theory and experiments in the limit of dense and saturated ensembles.
	
	In this paper, we show that an effective TLS can properly approximate the scattering properties of a MLS, with exact correspondence in certain conditions. To this end, we numerically solve  the optical Bloch equations (OBE) for a single MLS driven by a coherent field that originates either from a probe laser or from neighboring atoms via coherent scattering. We then fit the effective TLS model parameters to the coherent and total scattering rates obtained from the density matrix calculations. In this study, we detail the influence of the driving field polarization, stray magnetic fields and  incoherent resonant electromagnetic field background on the effective TLS parameters.
	
	Sec. \ref{sec:develop} introduces the relevant quantities discussed throughout this manuscript and derives the optical response of a TLS driven by a {coherent} field and an incoherent background. Sec. \ref{sec:MLS} details the calculation for the exact solution of a MLS and its comparison to an effective TLS. As an example, Sec. \ref{sec:MLS_TLS} quantitatively compares the  effective TLS that best corresponds to the closed transition of the D$_{2}$-line of $^{87}$Rb which is formed by the hyperfine states $F_{g}=2$ and $F_{e}=3$. The numerical simulations performed in Sec. \ref{sec:MLS_TLS} could be carried for any multi-level quantum system with a closed transition.

	\section{\label{sec:develop} Two-level system dynamics} 
	
	Our study begins with the scattering dynamics of two-level systems. We remind in Sec. \ref{subsec:scat_coh} the standard {expressions} \cite{Steck2019book,SteckRb872001} of the density matrix elements from which the coherent, incoherent and total scattering rates are derived. In Sec. \ref{subsec:scat_incoh}, we derive the {expressions} of the same quantities when the two-level systems TLS is driven by a coherent field and an incoherent field background.

	\subsection{\label{subsec:scat_coh} Scattering rate under coherent drive}
	The total scattering rate is an essential quantity describing the influence of the light field on the atom. It {gives} the number of photons emitted per unit of time \cite{Steck2019book}: $R^{(tot)}_{sca} =\Gamma \rho_{ee}$ where $\Gamma$ is the natural linewidth and $\rho_{ee}$ the excited state {population}. This total scattering rate can be decomposed in terms of a coherent scattering rate $R^{(coh)}_{sca}=\Gamma |{\rho}_{eg}|^2$ that represents scattering events that are temporally coherent with respect to the {driving} field and an incoherent scattering rate $R^{(inc)}_{sca}=R^{(tot)}_{sca}-R^{(coh)}_{sca}$ that, by energy conservation, is the difference between the two previous ones. Computing these rates requires deriving expressions for the density matrix population $\rho_{ee}$ and slowly varying coherence ${\rho}_{eg}$, both obtained by solving the steady-state regime of the OBE driven by a coherent field: 
	\begin{equation}
		\begin{aligned}
			{\rho}_{eg} = & \frac{-i}{\sqrt{2}} \frac{\sqrt{s_c'}}{ 1+s_c' } \frac{1+2i\delta}{\sqrt{1+4\delta^{2}}}, \\
			\rho_{ee} = & \frac{1}{2} \frac{s_c'}{1+s_c'},
			\label{eq:RhoeeRhoegSteady_copy}
		\end{aligned}
	\end{equation}
	where $s_c=2\Omega_{c}^{2}/\Gamma^{2}$ is the on-resonance saturation parameter, $s_{c}'=s_c/(1+4\delta^2)$ is the effective saturation parameter, $\Omega_{c}$ is the coherent Rabi frequency and $\delta=(\omega-\omega_{0})/\Gamma$ the normalized detuning between the laser frequency $\omega$ and the atomic transition $\omega_{0}$. 
	
	From Eqs. (\ref{eq:RhoeeRhoegSteady_copy}) the coherent and total scattering rates are given by:
	\begin{equation}
		\begin{aligned}
			R^{(coh)}_{sca} = & \frac{\Gamma}{{2}} \frac{{s_c'}}{ \left(1+s_c' \right)^2}, \\
			R^{(tot)}_{sca} = & \frac{\Gamma}{2} \frac{s_c'}{1+s_c'}.
			\label{eq:RhoeeRhoegSteady_Coherent_2level}
		\end{aligned}
	\end{equation}
	In the regime of weak saturation ($s_{c}\ll 1$), the atom response is linear in $s_{c}$ and temporally coherent. This is the regime of the linear dipole approximation that is convenient for coupled dipole simulations. In the opposite strong saturation regime ($s_{c}\gg 1$), the TLS can be saturated, and the temporally incoherent scattering dominates. This is the regime of the Mollow triplet where the incoherent scattering rate scales as ${s_{c}'^{2}}/{\left( 1+s_{c}'\right)^{2}}$.
	
	To move from an ideal TLS to the effective TLS, we follow the work of \cite{Gao1993} where the author computes analytically the scattering rates for the case of a $\pi$-polarization and shows that the saturation intensity is reduced by a factor $\alpha$. We then introduce an effective TLS ansatz under a saturating driving field and in perturbed conditions:
	\begin{equation}
		\begin{aligned}
			R^{(coh)}_{sca} = & \frac{\Gamma}{{2}} \frac{\beta}{\alpha} \frac{s_c'/\alpha}{ \left(1+s_c'/\alpha \right)^2}, \\
			R^{(tot)}_{sca} = & \frac{\Gamma}{2}   \frac{s_c'/\alpha}{1+s_c'/\alpha}.
			\label{eq:fit_functions}
		\end{aligned}
	\end{equation}
	Eqs. (\ref{eq:fit_functions}) correspond to an effective TLS with a corrected saturation parameter $s_c'/\alpha$. The factor $\beta/\alpha$ accounts for multi-level corrections (Sec. \ref{sec:MLS_TLS}) where $\alpha$ comes from a geometric factor due to the coupling strength of the transitions and $\beta$ is an amplitude factor of the coherent scattering field.
	
	\subsection{\label{subsec:scat_incoh} Scattering rate in coherent and incoherent drives}
	
	For a TLS, {coherences} between atomic states {are} driven by the field complex amplitude. Therefore, a temporally incoherent field (frequency broadband and/or temporally isotropic polarization) gives an average coherence of zero for averaging time longer than the field spectral width. Thus, only the intensity of the incoherent field affects the OBE by incoherently pumping the populations {at} a rate {of} $\Gamma s_{i}$/2 where $s_{i}$ is an effective saturation parameter for the incoherent intensity {\cite{Steck2019book}}. For a coherent field with Rabi frequency $\Omega_c$ and detuning $\delta$, and an incoherent intensity with saturation parameter $s_{i}$, one obtains the following OBE:

		\begin{align}
			\label{eq:Bloch_equations_2_level_incoherent}
			\frac{\partial{{\rho}}_{ge}}{{\partial}t} = & -(i{\delta\Gamma}+\frac{\Gamma}{2}){\rho}_{ge}+\frac{i\Omega_{c}}{2}( \rho_{gg} -\rho_{ee}), \\
			\frac{\partial{\rho}_{gg}}{{\partial}t} = & -\frac{s_{i}}{2}\Gamma \left(  \rho_{gg} -  \rho_{ee}  \right)  + \frac{i\Omega_{c}}{2}( {\rho}_{ge} -{\rho}_{eg}) +\Gamma\rho_{ee}. \nonumber
		\end{align}
	The steady-state solution of Eqs.(\ref{eq:Bloch_equations_2_level_incoherent}) is:
	
	\begin{equation}
		\begin{aligned}
			{\rho}_{eg} = & {\frac{-i}{\sqrt{2}} \frac{\sqrt{s_c'}}{\left( 1+s_c'+s_{i}\right)} \frac{1+2i\delta}{\sqrt{1+4\delta^2}}}, \\
			\rho_{ee} = & \frac{1}{2} \frac{s_c'+s_{i}}{1+s_c'+s_{i}}.
			\label{eq:RhoeeRhoegSteady_Incoherent}
		\end{aligned}
	\end{equation}
	Eqs. (\ref{eq:RhoeeRhoegSteady_Incoherent}) shows that populations can be transferred by both the coherent and incoherent light while coherences are driven only by the coherent field {but} damped in the saturation regime by both fields \textit{i.e.} coherences are reduced by the total saturation parameter $s_c'+s_{i}$.
	
	The above expressions of the density matrix elements of a TLS in an incoherent background (Eqs. \ref{eq:RhoeeRhoegSteady_Incoherent}) allow generalizing the effective TLS ansatz (Eqs. \ref{eq:fit_functions}) in:
	
	\begin{equation}
		\begin{aligned}
			R^{(coh)}_{sca} = & {\frac{\Gamma}{2}  \frac{\beta}{\alpha_{\rm eff}} \frac{{\frac{s_{c}'}{\alpha_{\rm eff}}}}{\left( 1+\frac{s_{c}'}{\alpha_{\rm eff}} \right)^{2}} },  \\
			R^{(tot)}_{sca} = & \frac{\Gamma}{2} \left( \frac{\frac{s_{c}'}{\alpha_{\rm eff}}}{1+\frac{s_{c}'}{\alpha_{\rm eff}}} +\frac{\frac{s_{i}}{\alpha_c}}{1+\frac{s_{i}}{\alpha_c}}\right),
			\label{eq:RhoeeRhoegSteady_Coherent_Incoherent_2level}
		\end{aligned}
	\end{equation}
	where the corrections to the scattering rates are:
	
	\begin{equation}
		\begin{aligned}
			\alpha_{\rm eff} =& \alpha \left(1+s_{i}\right), \\
			\alpha_c =& 1+s_c'.
			\label{eq:CorrectionScattRates}
		\end{aligned}
	\end{equation}

	\section{\label{sec:MLS} Multi-level system dynamics}
	
	In this section, we give the master equation that describes the density matrix evolution of a multi-level atomic system and the coherent and total scattering rates \cite{Steck2019book}.
	
	The master equation in the rotating frame at $\omega$ includes the hyperfine splitting Hamiltonian ${{H}}_{HF}$ coming from the atomic energy structure, the first order Zeeman magnetic shift Hamiltonian ${{H}}_{B}$ and the electric-dipole interaction Hamiltonian between the atom and the driving field:
	\begin{equation}
		\begin{aligned}
			\frac{d{\rho}}{dt} = &  - \frac{i}{\hbar} \left[ {{H}}_{HF} + {{H}}_{B} + {{H}}_{AF}  , {\rho} \right] \\ 
			& + \Gamma \left( \frac{2J_{e}+1}{2J_{g}+1} \right) \sum_{q} D[\Sigma_{q}] {\rho}, 
			\label{eq:master_equation}
		\end{aligned}
	\end{equation}
	where $\Sigma_{q}$ is the lowering operator for the polarization $q=m_{g}-m_{e}$, $D$ the Lindblad superoperator defined as $D[\Sigma_{q}]{\rho}=\Sigma_{q}{\rho}\Sigma_{q}^{\dagger}-1/2\left(\Sigma_{q}^{\dagger}\Sigma_{q}{\rho}+{\rho}\Sigma_{q}^{\dagger}\Sigma_{q}\right)$. The electric-dipole Hamiltonian is ${{H}}_{AF}=-\hat{\mathbf{d}}.{\mathbf{E}}$ where $\hat{\mathbf{d}}=-e. \hat{\mathbf{r}}$ is the dipole moment and $\mathbf{E}=E_0 \boldsymbol\epsilon  e^{i \omega t} + c.c.$.
	In this work, the master equation has been {traced} over the environment to obtain the OBE for the multi-level atom. Rewriting the density matrix components as a column vector $\boldsymbol\rho$, we obtain a set of linearly coupled equations:
	
	\begin{equation}
		\frac{d\boldsymbol\rho}{dt}=\bold{M}\boldsymbol\rho,
		\label{eq:set}
	\end{equation}
	where the matrix $\bold{M}$ contains the matrix elements for all interactions and are detailed in Eq. (\ref{eq:MEsingleModeField}) in the Appendix.
	
	The last interaction to be added to the master equation is an incoherent field background. The interaction between a broadband incoherent field such as a black-body radiation and a MLS has been previously studied \cite{Dodin2018,Dodin2016,Tscherbul2015}. Master equations enable one a detailed description of the system composed of an atom and a bath. We obtain the Lindblad equations via two simplifications. First the Born-Markov approximation assumes a weak coupling ($\Gamma \ll \omega_{0}$) and a vanishing bath memory time \cite{Dodin2018,Dodin2016,Tscherbul2015}. Secondly the secular approximation, where the coherences and the populations evolve independently, neglects interference effects (Zeeman coherences). Under these approximations, the master equation of an incoherently driven system simplifies to rate equations on the populations only. As detailed in the Appendix \ref{sec:appendix}, under the influence of an on resonance temporally incoherent field, the master equation of a coherently driven system is modified by including terms in the matrix $M$ to account for incoherent population transfer.
	
	For a multi-level atom, the coherent and total scattering rates in the steady-state are defined as:
	\begin{equation}
		\begin{aligned}
			R^{(coh)}_{sca} = & \Gamma \sum_{q} |\langle  \Sigma_{q}  \rangle|^{2},  \\
			R^{(tot)}_{sca} = & \Gamma \sum_{q}  \langle  \Sigma_{q}^{\dagger}\Sigma_{q}  \rangle .
			\label{eq:scattering_rates_multi_levels}
		\end{aligned}
	\end{equation}
	
	In our method, these rates are obtained by  solving the MLS master equation in the steady-state by setting ${d{\boldsymbol\rho}}/{dt}=0$ either with symbolic computations for pure polarizations to determine the exact analytical formulas or numerically otherwise. We checked that the results of both methods are totally consistent. From the density matrix solution, we compute the exact coherent and total scattering rates using Eq. (\ref{eq:scattering_rates_multi_levels}). The parameters $\alpha_{\rm eff}$ and $\beta_{\rm eff}$ are then obtained by fitting the exact rates with the effective TLS model from Eq. (\ref{eq:RhoeeRhoegSteady_Coherent_Incoherent_2level}). For simplicity, in the following, these parameters will be noted as $\alpha$ and $\beta$.
	
	\section{\label{sec:MLS_TLS} Effective TLS of the D$_2$-line of rubidium 87}
	
	To study the role of experimental imperfections such as polarization orientation, DC magnetic fields and incoherent background on the scattering rates, we restrict our model to multiple degenerate closed states. As an example, we choose to simulate our model on all Zeeman states of the transition $\ket{F_{g}=2} \rightarrow \ket{F_{e}=3}$ of the $^{87}$Rb D$\phantom{}_{2}$-line and restrict ourselves to situations where the power broadening is much smaller than the hyperfine energy splitting. Also, the ratio of scattering rates between the $\ket{F_{e}=3}$ and $\ket{F_{e}=2}$ states (44 $\Gamma$ detuned from $\ket{F_{e}=3}$) is about 1000 for $s_c=60$ on the closed transition. As a result, the transition from $\ket{F_{g}=2}$ to $\ket{F_{e}=3}$ is considered to be closed in the range $s_c \in [0.1,30]$. In the following, we therefore neglect the residual coupling to other hyperfine excited and ground states. In the steady-state regime, these coupling would lead to depumping out of the considered transition. Our study is therefore valid only before depumping occurs and is robust for closed transitions.

	The role of the driving field polarization (Sec. \ref{subsec:scat_coh_multi_sigma_pi}), DC magnetic field (Sec. \ref{subsec:scat_coh_multi_B}) and isotropic incoherent field (Sec. \ref{subsec:scat_incoh_multi}) are studied independently. The quantization axis is taken as $\boldsymbol\epsilon_{z}$.

	\subsection{Role of polarization: $\sigma_{\pm}$, $\pi$ and elliptical\label{subsec:scat_coh_multi_sigma_pi}}

	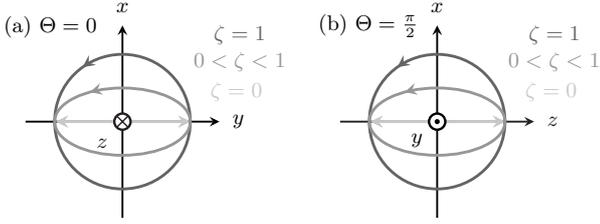
\begin{figure}[t]
		\begin{center}
			\begin{tabular}{ll}
				\resizebox{4cm}{!}{  
					\begin{tikzpicture}[
						scale=1,
						arrow_axis/.style={thick,->,shorten >=2pt,shorten <=2pt,>=stealth},
						linear/.style={very thick,-,shorten >=2pt,shorten <=2pt,>=stealth,linear_gray,decoration={markings, mark=at position 0.1 with {\arrow{<}}},postaction={decorate},decoration={markings, mark=at position 1 with {\arrow{>}}},postaction={decorate}},  ]
						\node[fill=white,above right=0.1cm of {(-2,1)}, outer sep=2pt,fill=white] {(a) $\Theta=0$};
						
						\draw[arrow_axis] (-1.5cm,0cm) -- (1.5cm,0cm) node[right] {$y$};
						\draw[arrow_axis] (0cm,-1.5cm) -- (0cm,1.5cm) node[above] {$x$};
						
						\draw[linear] (-1cm,0cm) -- (1cm,0cm) node[fill=white,above right=0.1cm of {(1.1,0.1)}] {$\zeta=0$};
						
						\draw[very thick,-,shorten >=2pt,shorten <=2pt,>=stealth,circle_gray,decoration={markings, mark=at position 0.35 with {\arrow{>}}},postaction={decorate}] (0,0) circle (1) ;
						\node[circle_gray,fill=white,above right=0.1cm of {(1.1,0.9)}, outer sep=2pt,fill=white] {$\zeta=1$};
						
						\draw[very thick,-,shorten >=2pt,shorten <=2pt,>=stealth,ellipse_gray,decoration={markings, mark=at position 0.35 with {\arrow{>}}},postaction={decorate}] (0,0) ellipse (1cm and 0.5cm);
						\node[ellipse_gray,fill=white,above right=0.1cm of {(0.8,0.5)}, outer sep=2pt,fill=white] {$0<\zeta<1$};
						
						\filldraw[Black,fill=white,line width=1pt](0,0)circle(.12cm);
						\draw[Black,line width=.6pt] (0,0)
						+(-135:.12cm) -- +(45:.12cm)
						+(-45:.12cm) -- +(135:.12cm);  
						\draw[Black](-0.3,-0.50)node[above]{$z$};
						
				\end{tikzpicture}} &
				\resizebox{4cm}{!}{  
					\begin{tikzpicture}[
						scale=1,
						arrow_axis/.style={thick,->,shorten >=2pt,shorten <=2pt,>=stealth},
						linear/.style={very thick,-,shorten >=2pt,shorten <=2pt,>=stealth,linear_gray,decoration={markings, mark=at position 0.1 with {\arrow{<}}},postaction={decorate},decoration={markings, mark=at position 1 with {\arrow{>}}},postaction={decorate}},  ]
						\node[fill=white,above right=0.1cm of {(-2,1)}, outer sep=2pt,fill=white] {(b) $\Theta=\frac{\pi}{2}$};
						
						\draw[Black,arrow_axis] (-1.5cm,0cm) -- (1.5cm,0cm) node[right] {$z$};
						\draw[arrow_axis] (0cm,-1.5cm) -- (0cm,1.5cm) node[above] {$x$};
						
						\draw[linear] (-1cm,0cm) -- (1cm,0cm) node[fill=white,above right=0.1cm of {(1.1,0.1)}] {$\zeta=0$};
						
						\draw[very thick,-,shorten >=2pt,shorten <=2pt,>=stealth,circle_gray,decoration={markings, mark=at position 0.35 with {\arrow{>}}},postaction={decorate}] (0,0) circle (1) ;
						\node[circle_gray,fill=white,above right=0.1cm of {(1.1,0.9)}, outer sep=2pt,fill=white] {$\zeta=1$};
						
						\draw[very thick,-,shorten >=2pt,shorten <=2pt,>=stealth,ellipse_gray,decoration={markings, mark=at position 0.35 with {\arrow{>}}},postaction={decorate}] (0,0) ellipse (1cm and 0.5cm);
						\node[ellipse_gray,fill=white,above right=0.1cm of {(0.8,0.5)}, outer sep=2pt,fill=white] {$0<\zeta<1$};
						
						\filldraw[fill=white,line width=1pt](0,0)circle(.12cm);
						\filldraw[fill,line width=1pt](0,0)circle(.02cm);
						\draw(-0.3,-0.50)node[above]{$y$};
						
				\end{tikzpicture}}
			\end{tabular}
			\caption{Two standard polarization cases and their {dependence on} ellipticity $\zeta$ and {quantization} axis (${z}$): a) For $\Theta=0$, the polarization lies in the (x,y) plane and is circular for $\zeta=1$ with a perpendicular quantization axis, b) For $\Theta=\pi/2$, the polarization lies in the (x,z) plane and is linear for $\zeta=0$ with a parallel quantization axis.}
			\label{fig:polar_cases_1_2}
		\end{center}
	\end{figure}

	In this section, only the coherent drive field polarization is being {changed} at zero magnetic field and zero incoherent field. We use the spherical basis with respect to the {Cartesian} as follows:
		\begin{align}
			\boldsymbol\epsilon_{\pm} = & \mp \frac{\left( \boldsymbol\epsilon_{x} \pm i \boldsymbol\epsilon_{y}\right)}{\sqrt{2}}, \\
			\boldsymbol\epsilon_{0} = & \boldsymbol\epsilon_{z}. 
			\label{eq:spherical_components_Cartesian_components}
		\end{align}
	This polarization ${\boldsymbol\epsilon}=({\epsilon}_{-},{\epsilon}_{+},{\epsilon}_{0})$ is parametrized in the spherical basis $(\boldsymbol\epsilon_{-},\boldsymbol\epsilon_{+},\boldsymbol\epsilon_{0})$ by an ellipticity $\zeta=E_{0x}/E_{0y}$ and $\pi$ polarization projection angle $\Theta$ (Fig. \ref{fig:polar_cases_1_2}) as:
	\begin{equation}
		\begin{aligned}
			{\epsilon}_{\pm} = & \mp \frac{1}{\sqrt{2}\norm{{\boldsymbol\epsilon}}} ( \zeta \mp \cos\Theta ), \\
			{\epsilon}_{0} = & \frac{i}{\norm{{\boldsymbol\epsilon}}} \sin\Theta,
			\label{eq:define_spherical_components}
		\end{aligned}
	\end{equation}
	where $\norm{{\boldsymbol\epsilon}}=\sqrt{1+\zeta^2}$.
	Using the TLS ansatz of Eq. (\ref{eq:fit_functions}), we evaluate the parameters $\alpha\equiv\alpha_{\boldsymbol\epsilon}$ and $\beta\equiv\beta_{\boldsymbol\epsilon}$ that best {match} the exact scattering rates. For a given $\zeta$, both scattering rates $R^{(coh)}_{sca},
	R^{(tot)}_{sca} $ are computed numerically as a function of the {coherent} saturation parameter $s_c$. $\alpha_{\boldsymbol\epsilon}$ and $\beta_{\boldsymbol\epsilon}$ are the best fitting parameter for $s_c \in [0.1,30]$.\\ 
    In the limit case of a $\sigma_-$ circular polarization $\boldsymbol\epsilon=(1,0,0)$ parametrized by $(\zeta,\Theta)=(1,0)$, the atom is pumped in a perfect two-level cycling transition and is expected to reach the maximal scattering cross section $\sigma_{0}={3\lambda^2}/{2\pi}$ with $\alpha_{\sigma}=\beta_{\sigma}=1$. In the opposite limit of a $\pi$ linear polarization $\boldsymbol\epsilon=(0,0,1)$, the populations and coherences have been computed analytically \cite{Gao1993}. The system formed by five $\pi$-transitions for $\ket{F_{g}=2}$ to $\ket{F_{e}=3}$ is equivalent to a TLS with a reduced cross section for which $\alpha_{\pi}=461/252=1.829$ and $\beta_{\pi}=1$.

\begin{figure}[t]
	\begin{center}
		\includegraphics[width = 0.5\textwidth]{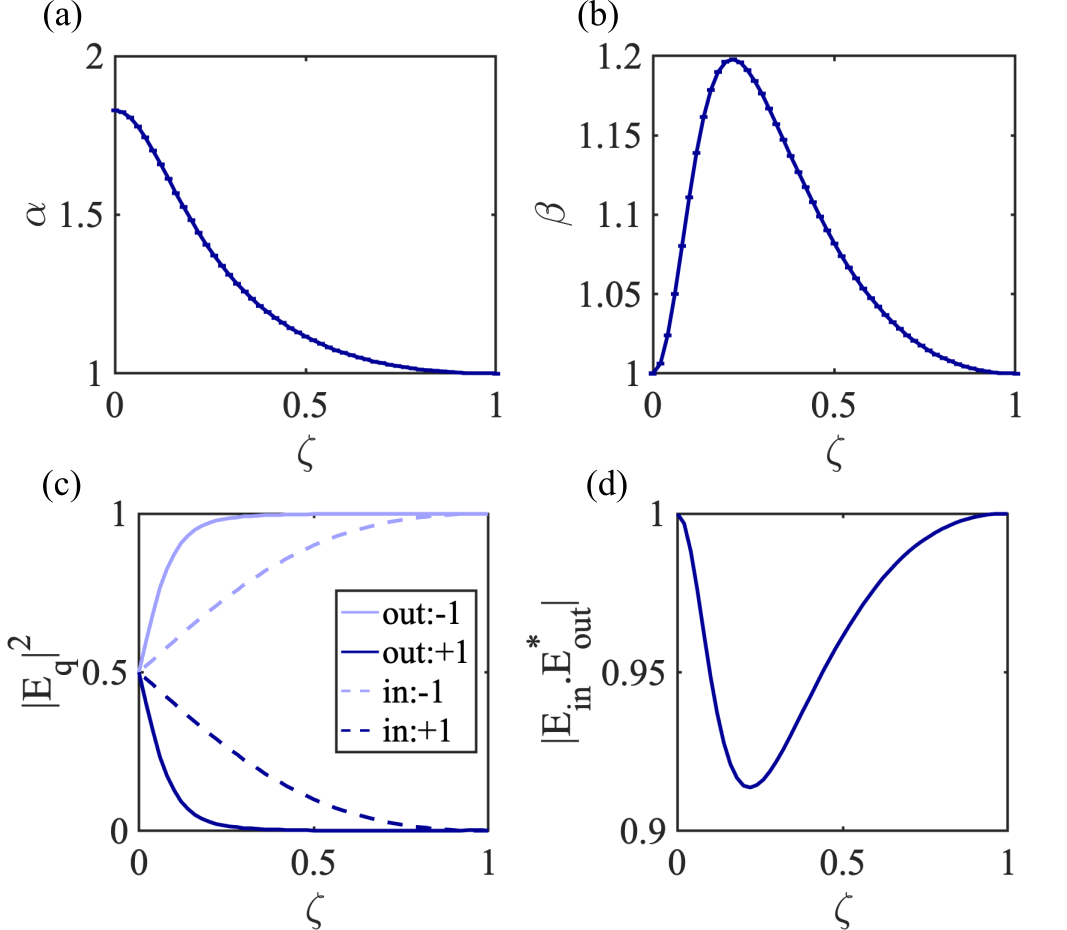} 
		\caption{a) $\alpha$ and b) $\beta$ as a function of the ellipticity $\zeta$ for a polarization in the (x,y) plane ($\Theta=0$), c) input and output intensities per polarization, d) scalar product of the input and output fields.}
		\label{fig:alpha_beta_function_elli_at_B0_0}
	\end{center}
\end{figure}
In Fig. \ref{fig:alpha_beta_function_elli_at_B0_0}, intermediate ellipticities are obtained by varying $\zeta$ from 1 (circular) to $0$ (linear) at $\Theta=0$. The values of $\alpha_{\boldsymbol\epsilon}$ and $\beta_{\boldsymbol\epsilon}$ are {exact} (error bars go to 0) which means that the scattering rates are {also} exactly described by Eq. (\ref{eq:fit_functions}). It does not necessarily mean that this situation is exactly equivalent to a TLS, that is a scalar scattering problem, since the MLS is vectorial.
The circular and linear {polarizations} {yield} the expected {values} $\alpha_{\sigma}=1$ and $\alpha_{\pi}=1.829$. Note that the linear polarization is at {a} \ang{45} {angle} of the $x$ and $y$ axis. In absence of magnetic field bias, the electric field sets the quantization axis. The same curves would be obtained for $\Theta=\pi/2$.
An imperfect polarization as could occur in experiments induces little changes for the circular polarization ($\zeta=0$) -below $10\%$ variation on the parameters- even up to $\zeta=0.5$, while it has more effect for a linear polarization. This is due to optical pumping in the closed transition that protects the atomic state.
For the same reason, the polarization of the radiated field differs from the input polarization at maximum by 8$\%$ for an ellipticity about $\zeta=0.2$.

\subsection{Role of a DC magnetic field\label{subsec:scat_coh_multi_B}}

Under a constant magnetic field, magneto-optical effects occur. We refer to the review \cite{Budker2002} for more detail. The Faraday effect, for example,
results in the optical rotation and ellipticity change of the output scattering. The scattering process is therefore a truly vectorial problem that cannot be exactly mapped onto the TLS solution due to the output polarization.

	\begin{figure}[h]
	\begin{center}
		\includegraphics[width=0.5\textwidth]{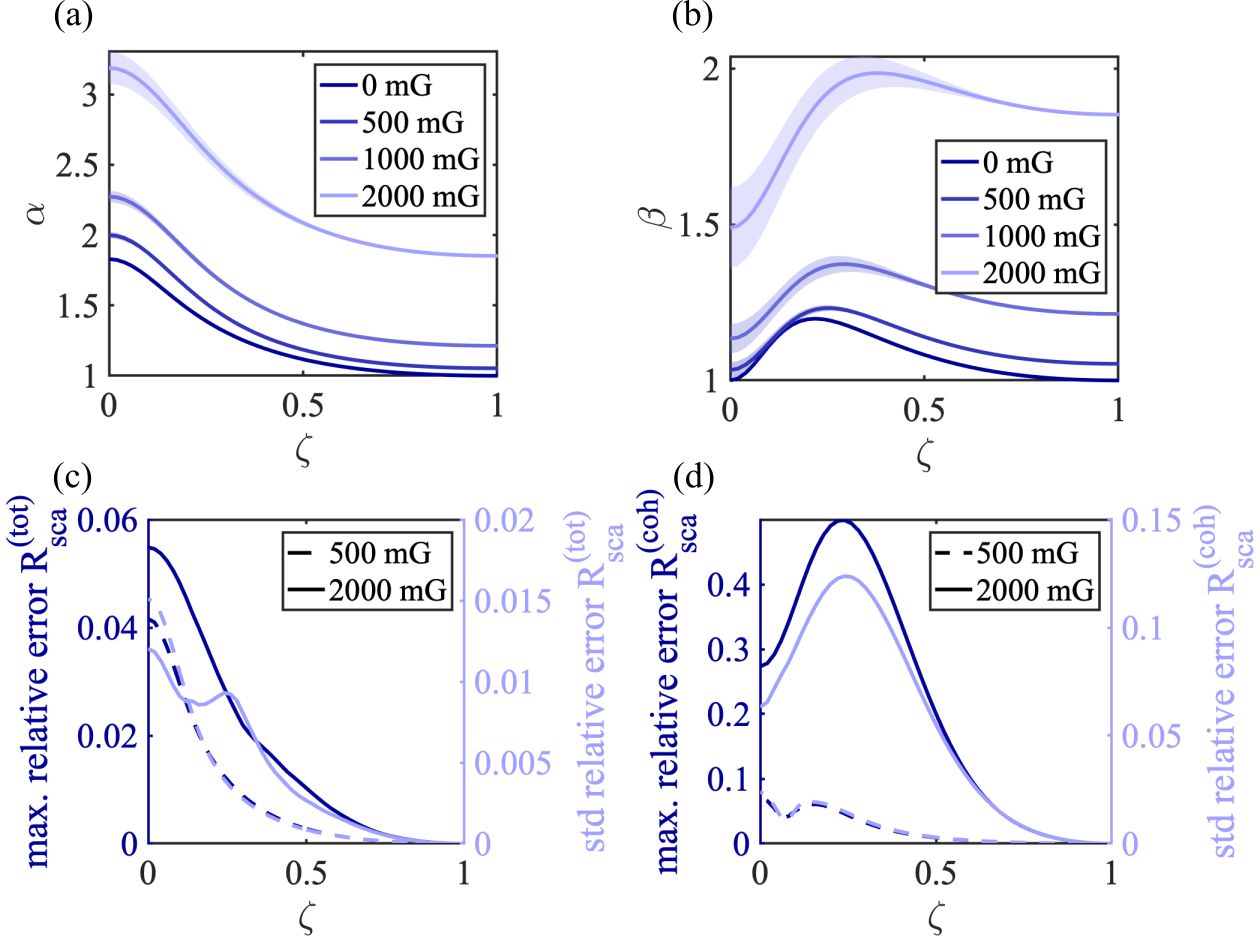} 
		\caption{a) $\alpha$ and b) $\beta$ as a function of the ellipticity $\zeta$ in the case of $\Theta=0$. $\zeta=1$ is circular and $\zeta=0$ is linear along $\boldsymbol\epsilon_{x}$ and does not correspond to a $\pi$ polarization. For any ellipticity, $\mathbf{B}$ is perpendicular to the electric field. The maximum and the {standard deviation} of the relative error between the approximated TLS solution and the exact calculation of the c) excited state population and d) density matrix coherence.}
		\label{fig:alpha_beta_function_elli_all_B0_pol1}
	\end{center}
\end{figure}
In this section, we focus the analysis on the TLS parameters $\alpha$ and $\beta$ that best mimic the scattering rate amplitude when the Zeeman degeneracy is lifted by a magnetic field bias such that $\mathbf{B}=B\boldsymbol\epsilon_{z}$. The driving field frequency is kept constant and is equal to the unshifted transition. The magnetic Zeeman shift between two states $\ket{F_{e},m_{e}}$ and $\ket{F_{g},m_{g}}$ is $\Delta(B)= \mu_{b}B/\hbar \left(  g_{F_{e}} m_{g}  -  g_{F_{g}} m_{g}  \right)$ where $g_{F_{e/g}}$ are the Land\'{e} factors and $\mu_b$ is the Bohr magneton (see Appendix \ref{sec:appendix}). {The strength of magnetic field considered in this study are up to $2$G which is well below the restriction to first order Zeeman perturbation and results in a frequency shift smaller than $\Gamma$. The state $\ket{F_{g}=2}$ (resp. $\ket{F_{e}=2}$) has a frequency sensibility to magnetic shift of 0.12$\Gamma$/G (resp. 0.15$\Gamma$/G).
	
	In a $\sigma_{-}$ polarization case, $\alpha=\beta=\alpha_{\sigma}(1+4\left(\delta_{\omega}+\delta_{B}\right)^{2})$ where $\delta_{B}=\mu_{b}B/\hbar\Gamma$. It simply corresponds to a TLS probed off-resonantly due to the Zeeman shift and the driving field detuning. Due to optical pumping, only the two states $\ket{F_{g}=2,m_{g}=-2}$ and $\ket{F_{e}=3,m_{e}=-3}$ are occupied in the steady-state. The atomic response is the one of a TLS and the results of Eq. (\ref{eq:RhoeeRhoegSteady_Coherent_2level}) are exactly recovered. In this situation, the Zeeman shift can be experimentally compensated by the driving field detuning.
	Interestingly, for a linear polarization aligned with a magnetic bias ($\pi$ polarization), the scattering rates are also exactly given by the effective TLS Eqs. (\ref{eq:RhoeeRhoegSteady_Coherent_2level}) with  $\alpha=\alpha_{\pi}\left(1+4\delta_{B}^{2}\frac{41}{1008\alpha_{\pi}}+4\delta_{\omega}^{2}\right)$ and $\beta=1$. These expressions can be derived by solving the linear system of Eqs. (\ref{eq:set}) containing all $\pi-$transitions. In this situation, the sensitivity of the scattering rates to the detuning is reduced by a factor $\frac{41}{1008\alpha_{\pi}}\approx 0.02$ with respect to the $\sigma$ polarization case.
	As a result, for pure $\sigma$ or $\pi$ polarizations, $\alpha$ and $\beta$ have well known lower and upper limits. 
	
	\begin{figure}[t]
		\begin{center}
			\includegraphics[width=0.5\textwidth]{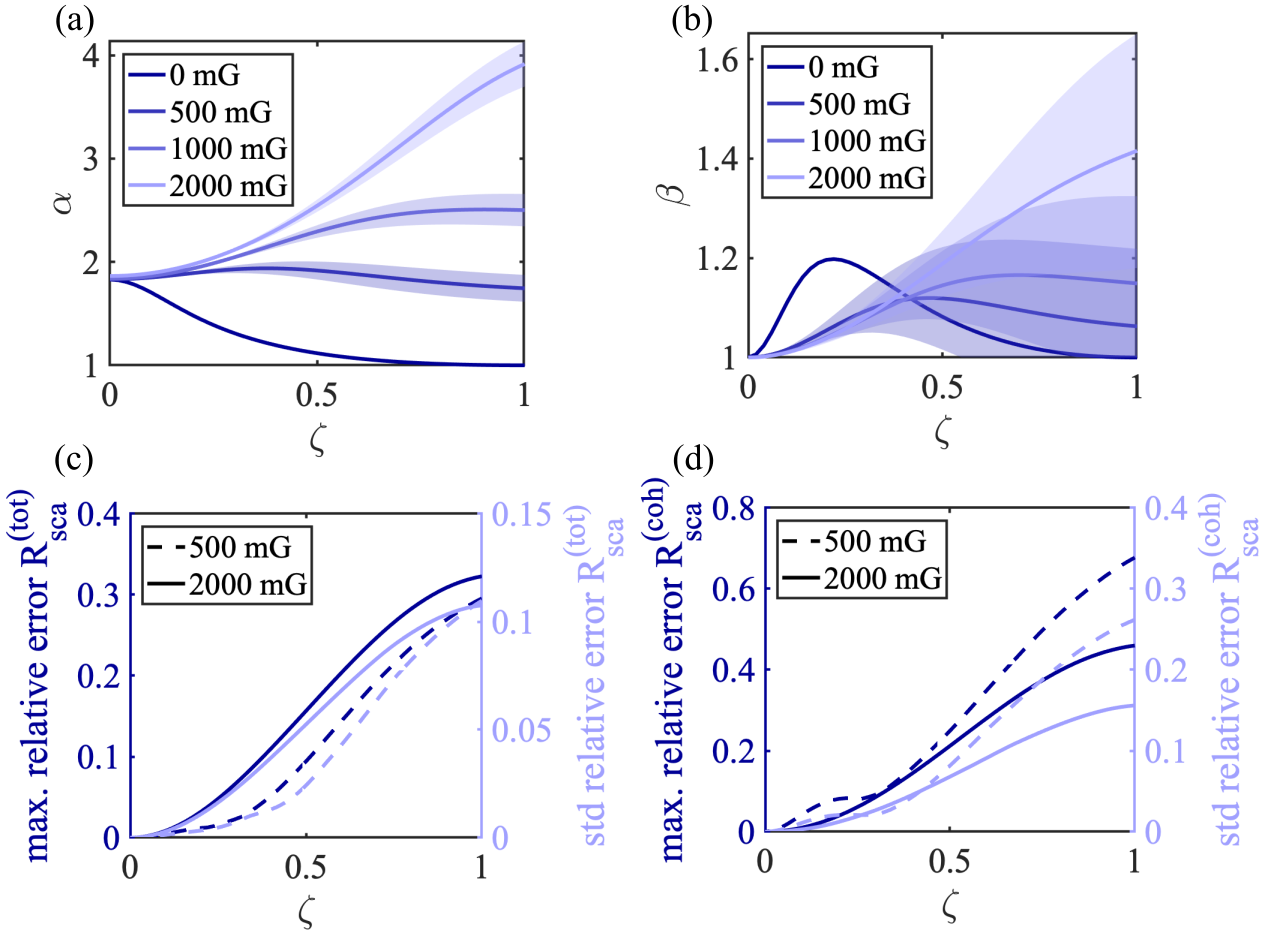} 
			\caption{a) $\alpha$ and b) $\beta$ as a function of $\epsilon$ in the case of $\Theta=\pi/2$. $\zeta=1$ is circular and $\zeta=0$ is linear along $\boldsymbol\epsilon_{z}$ and does correspond to a $\pi$ polarization. The maximum and the {standard deviation} of the relative error between the approximated TLS solution and the exact calculation of the c) excited state population and d) density matrix coherence.}
			\label{fig:alpha_beta_function_elli_all_B0_pol2}
		\end{center}
	\end{figure}
	We consider now deviations from these ideal cases by studying the influence of the ellipticity $\zeta$ in the $(x,y)$ plane corresponding to $\Theta=0$, and in the $(x,z)$ plane corresponding to $\Theta=\pi/2$. $\alpha$ and $\beta$ from Eq. (\ref{eq:RhoeeRhoegSteady_Coherent_Incoherent_2level}) are together fitted from the scattering rates obtained by numerically solving Eq. (\ref{eq:set}) and plotted with error bars within $95\%$ confidence interval of the fits parameters.
	
	As expected, Fig. \ref{fig:alpha_beta_function_elli_all_B0_pol1} ($\Theta=0$) shows that the scattering rates are sensitive to the polarization. The growing error bars for an increased ellipticity indicate a growing deviation from the effective TLS behavior. The offset between the curves is merely due to the Zeeman shift. The relative error between the simulated scattering rate and the model of Eq. (\ref{eq:fit_functions}) is plotted in terms of the maximum and the standard deviation of the relative error over the range of the saturation parameter $s_c$ used for the fitting. It shows that the model has a maximum relative error which does not exceed $6\%$ for the total scattering rate, which happens only for an ellipticity of 1. The coherent scattering rate is more sensitive to the ellipticity with up to $50\%$ relative error. Nevertheless, for quasi-circular polarizations ($\zeta\approx1$) as typically used in $\sigma$ absorption imaging, the MLS reduces to a TLS.
	
	Fig.  \ref{fig:alpha_beta_function_elli_all_B0_pol2} presents the influence of the ellipticity in the ($x,z$) plane. For $\zeta=0$, corresponding to a $\pi$ polarization, the TLS is exact with little dependence on magnetic field. As the ellipticity is increased, this TLS behavior becomes less accurate with  increasing error bars on the fitting parameters. Consequently, $\pi$ polarization imaging will be less accurate to define the absorption cross section and the atom numbers.
	
	\subsection{Role of an incoherent drive\label{subsec:scat_incoh_multi}}
	The scattering cross section is also modified in the presence of an incoherent field. It could be for example generated by a thermal lamp, or by the temporally incoherent response of the surrounding atomic gas to the coherent excitation. Here, we consider the influence of a partially polarized incoherent field on the coherent and incoherent scattering response of a MLS probed by a coherent $\sigma_-$ polarized light under zero magnetic field offset.
	Electromagnetic fields are here considered as temporally incoherent if their coherence time is smaller than the Rabi period, meaning that the density matrix coherences are zero on average while the excited state populations are non-zero. An incoherent field yields optical pumping without coherence.
	
	We consider a  partially $\sigma_-$ polarized  incoherent field that could for example be generated by the incoherent scattering of a $\sigma_-$ polarized coherent field. This incoherent field is parametrized by a polarization degree $r\in[0,1]$ in the form $s_{i}^{\sigma}=r s_{i}$ and $s_{i}^{\rm iso}=(1-r) s_{i}$ where the $s_{i}^{\sigma}$/$s_{i}^{\rm iso}$/$s_{i}$ are respectively the $\sigma_-$ polarized / isotropic / total saturation intensities expressed relatively to the saturation intensity of $\sigma_-$ polarized light. With such definition, $r=1$ describes a $\sigma_-$ polarized incoherent field while $r=0$ describes an isotropic incoherent field.
	
	In the limit $r=1$, both the coherent and incoherent fields are $\sigma_-$ polarized. The atomic population are pumped in the closed transition and the situation is exactly described as in Sec. \ref{subsec:scat_incoh} with scattering rates given by Eq. (\ref{eq:RhoeeRhoegSteady_Coherent_Incoherent_2level}) where $\delta_{}=0$ and $\alpha=1$.
	In the opposite limit $r=0$, a purely isotropic incoherent field drive will redistribute the ground state population. In the absence of the coherent field drive ($s_c=0$), the populations are equally redistributed among the Zeeman states. Interestingly, we notice that for large saturation parameters $s_{i}$, the total population in the excited state can be higher than $1/2$. It is indeed bounded by the number of excited Zeeman states over the total number of states which is  $7/12$ for the considered transition of $^{87}$Rb. The total excited state population $\rho_{ee}$ in the absence of coherent field $s_c=0$ can be analytically derived from the master equation (Appendix \ref{sec:appendix}) and is given by:
	
	\begin{equation}
		\begin{aligned}
			\rho_{ee} = & \frac{7}{12} \frac{s_{i}/\frac{30}{12}}{1+s_{i}/\frac{30}{12}}.
			\label{eq:incoherent_tot_pop}
		\end{aligned}
	\end{equation}
	In the low saturation regime \textit{i.e.} $s_{i}\ll1$, Eq. (\ref{eq:incoherent_tot_pop}) reduces to $\rho_{ee}=\frac{1}{2}\frac{s_{i}}{\alpha_{\rm iso}}$ where $\alpha_{\rm iso}=\frac{15}{7}$ is the reduction of the cross section for an isotropic and incoherent field. 
	
	We consider now the general case of a MLS driven simultaneously by a partially polarized incoherent field ($r\in [0, 1]$) and by a coherent drive $s_c$. In the Appendix \ref{sec:appendix}, the master equation given by Eq. (\ref{eq:MEsingleModeField}) is expressed in the form of Eq. (\ref{eq:ME_incoherent_terms}) and the steady-state solutions of all elements of the density matrix are obtained by an algebraic solver. The algebraic solution has the form of a ratio of polynomials containing few 100 terms. 
	This exact algebraic solution of the scattering rates is compared to a phenomenological model of a TLS:
	\begin{eqnarray}
			R_{sca}^{(coh)}/\Gamma &= &  \frac{\beta}{2 \alpha}  \frac{s_{c}/\alpha}{\left(1+s_{i}^{\rm iso}/\frac{30}{12}+s_{i}^\sigma/\alpha+s_c/\alpha\right)^2},\nonumber\\
			R_{sca}^{(tot)}/\Gamma &= & \eta(r,s_i,s_c) + \frac{1}{2}  \frac{ s_{c}/\alpha}{1+s_{i}^{\rm iso}/\frac{30}{12}+s_{i}^\sigma/\alpha+s_c/\alpha},\nonumber\\
			\eta(r,s_i,s_c)& =& \frac{\eta_0-\eta_\infty}{1+\frac{s_c}{\alpha \left( 1+s_i^{\rm iso}/\frac{30}{12}+ s_{i}^\sigma/\alpha\right)}}+\eta_\infty,
			\label{eq:fit_multi-level_ScaRates}
	\end{eqnarray}
	where $\alpha$, $\beta$, $\eta_0$ and $\eta_\infty$ are fitting parameters. To minimize the number of fitted parameters, the saturation intensity correction of the isotropic incoherent field  was fixed to $30/12$. In this model, $\alpha$ quantifies  the reduction of the coherent absorption cross section, $\beta<1$ quantifies the reduction of coherent field emission with respect to a perfect TLS. $\eta(r,s_i,s_c)$ is the part of the total excited state population induced by the incoherent field drive. This contribution of the excited state population includes states which are coupled to all three fields (e.g. $\ket{F_e=3,m_F=-3,-2,-1,0,1}$) and states coupled only to the isotropic polarization (e.g. $\ket{F_e=3,m_F=2,3}$). This excited state population $\eta(r,s_i,s_c)$ has a complex dynamic that depends on the polarization ratio of the incoherent field $r$, the value of the incoherent $s_i$ and coherent $s_c$ saturation intensities.  For a given couple of parameters $r$ and $s_i$, we observe that this contribution varies monotonically from $\eta_0$ in absence of coherent drive and saturates at $\eta_\infty$ for large coherent drive $s_c\gg s_i$. The model given in Eq. (\ref{eq:fit_multi-level_ScaRates}) reproduces the saturation behavior with a crossover at $s_c/\alpha = 1+s_i^{\rm iso}/\frac{30}{12}+ s_{i}^\sigma/\alpha$.\\
	As an example, the total scattering rate and coherent scattering rate are given as a function of $s_c$ for $r=0.5$ and $s_i=5$ in Fig. \ref{fig:ex_r0p5_incoherent}. We observe that the equivalent TLS given by Eqs. (\ref{eq:fit_multi-level_ScaRates}) describes well the dynamics with a standard deviation of the relative error below {$1\%$ on this example}. The excited state population offset at $s_c=0$ is due to the incoherent drive.
	

	\begin{figure}[t]
	   \begin{center}
		\includegraphics[width = 0.5\textwidth]{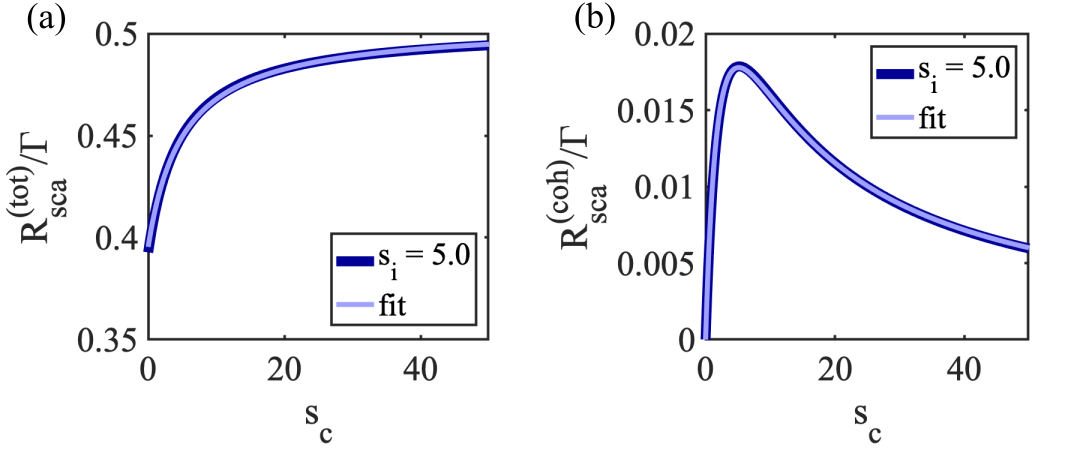} 
			\caption{a) Variations of the total excited state population and b) $\sigma_-$-polarized coherences with $r=0.5$ and $s_{i}=5$ in blue and the fits in red using Eq. ($\ref{eq:fit_multi-level_ScaRates}$).}
			\label{fig:ex_r0p5_incoherent}
		\end{center}
	\end{figure}
	We compare the TLS model and the exact solution by fitting the parameters $\alpha(s_i,r)$, $\beta(s_i,r)$ $\eta_0(s_i,r)$ and $\eta_\infty(s_i,r)$. The fits are realized at a fixed incoherent intensity $s_i \in [0,10]$ and polarization degree $r\in [0,1]$ and for a  coherent intensity $s_c$ varying between 0 and $10 s_i$. In this parameter range ($s_c\leq 100$) the closed transition approximation holds at least for 500 scattering events. The results are shown in Fig. \ref{fig:alpha_beta_gamma_eta_incoherent}. As a comparison, the expected parameters for an exact TLS are $\alpha=1, \beta=1$ and $\eta_0=\eta_\infty$ given by the second term in Eq. (\ref{eq:RhoeeRhoegSteady_Coherent_Incoherent_2level}) which are indeed the exact solutions found for a polarized incoherent field ($r=1$). For $r=1$, the MLS is optically pumped in an exact TLS. As the incoherent drive polarization is randomized $r\rightarrow0$, the fitted parameters slightly deviate from these initial values with a complex behavior that is mostly driven  by optical pumping mechanisms. In absence of coherent drive, the excited state population $\eta_0$ increases as a function of the incoherent saturation intensity. For an isotropic incoherent drive, this excited state population can even exceed 1/2 (\textit{cf}. Eq. (\ref{eq:incoherent_tot_pop})) as all 12 states become equally populated. In our model, the coherent scattering rate only depends on $\alpha$ and $\beta$. For large and unpolarized incoherent drive,  the coherent scattering rate can be reduced by up to a factor 1/2 with respect to the pure TLS. This reduction is essentially due to  Clebsch-Gordon coefficients entering in the calculation of $\beta$. As checked numerically, the polarization of the coherently scattered field is exactly aligned with the input field ($\sigma$-polarized). This was expected given that the incoherent field cannot drive coherence between the Zeeman sub-levels and therefore does not alter the scattered polarization (Eq. (\ref{eq:optical_coherences_eg}) in the Appendix).
	
		\begin{figure}[t]
		\begin{center}
			\includegraphics[width=0.5\textwidth]{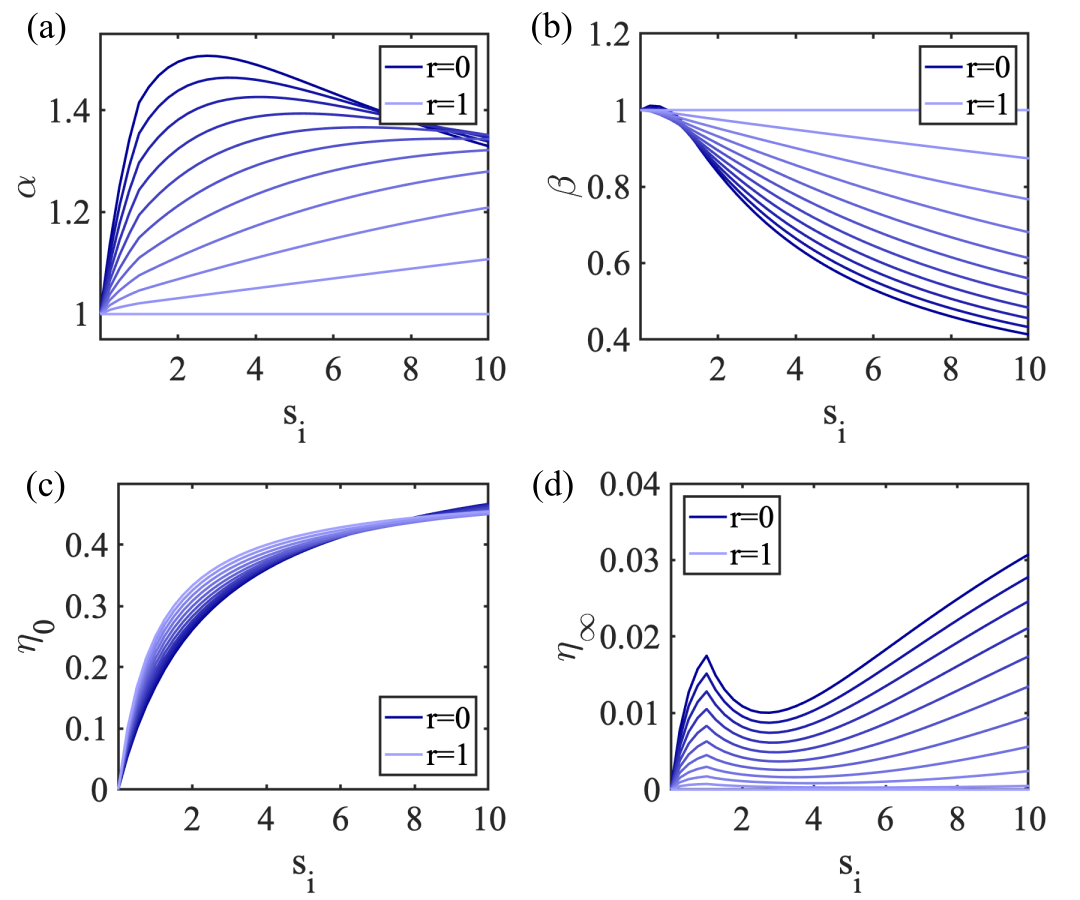} 
			\caption{Model parameters a) $\alpha$, b) $\beta$, c) $\eta_0$ and d) $\eta_\infty$ as a function of the incoherent intensity $s_i$ for polarization degrees $r\in [0,1]$ by step of 0.1. The shaded area represents the $95\%$ confidence interval of the fits and is very small on the plots. In c), the limit case $r=0$ and $r=1$ exactly coincide with the analytical solutions given in Eq. (\ref{eq:incoherent_tot_pop}) and the second term of Eq. (\ref{eq:RhoeeRhoegSteady_Coherent_Incoherent_2level}) respectively.}
			\label{fig:alpha_beta_gamma_eta_incoherent}
		\end{center}
	\end{figure}

	
	\section{Discussion} \label{sec:discussion}

	The combined coherent and incoherent response of atoms in the saturated regime has a strong impact on the interpretation of fluorescence and absorption imaging of ensembles. Experimental imperfections are the common explanations \cite{Reinaudi2007} for {the reduction of} the atomic cross section. With this argument, the absolute determination of atom numbers is subject to a precise calibration of the experimental conditions which is often questioned. In Fig. \ref{fig:alpha_beta_function_elli_at_B0_0}, we have shown that the effective TLS description of a multi-level atom is very robust to polarization imperfections especially for the case of $\sigma$ polarized light. We attribute this robustness to optical pumping mechanisms that protects the stretched-state hyperfine transition ($|F_g=2, m_F=-2\rangle$ to $|F_e=3, m_F=-3\rangle$ in $\sigma_-$). This pumping was additionally observed in the steady-state solutions of the OBE via a strong imbalance of the repartition of the population that favored the $|F_g=2, m_F=-2\rangle$ state even for $\zeta\in [0.5,1]$. This optical pumping protection of $\sigma$ transitions is not specific to the imaging transition of the D$\phantom{}_2$ line of Rubidium and will be applicable to any Zeeman degenerated closed transition. In the absence of Zeeman degeneracy, it was additionally shown that MLS behaves exactly as an effective TLS for any drive polarization. As expected for $^{87}$Rb, the cross section reduction factor $\alpha$ varies from 1 ($\sigma$) to 1.829 ($\pi$) depending on the driving field polarization.
	
	On the other hand, stray magnetic field  strongly impact the calibration of the scattering cross section of a $\sigma$ polarized probe (Fig. \ref{fig:alpha_beta_function_elli_all_B0_pol1}(a)). Nevertheless, this correction which is solely given by detuning of the coherent drive from the extreme TL transition can be exactly compensated for by tuning the drive frequency on resonance with the extreme TL transition in experiments. It is therefore not a concern for cross section calibrations. In addition, due to symmetry of the first order Zeeman splitting, the cross section correction of a $\pi$ polarized probe is independent of the stray magnetic field amplitude if it is aligned along the linear polarization axis.
	
	As mentioned earlier, in the presence of a stray magnetic field, light scattering is a vectorial process, and the coherent scattered field polarization is not aligned with the drive. Nevertheless, in the specific case of $\sigma$ and $\pi$ polarized light with a magnetic bias well-defined with respect to light polarization, both atom-light and magnetic interaction are diagonalized in the same basis leading to an aligned output field. The scattering process is therefore scalar in these two situations with additional robustness to imperfections for the $\sigma$ polarized drive (Fig. \ref{fig:alpha_beta_function_elli_at_B0_0}(d)). Polarization and stray magnetic field imperfections have therefore little influence on the reduction of the cross section.
	
	On the other hand, we have observed that in the presence of an incoherent background that would mimic {the temporally and spatially incoherent scattering from other atoms in the ensemble}, the cross section is notably reduced which results in systematics errors on absorption imaging measurements of atom number in the saturation regime and/or at large optical thickness \cite{Veyron21,Corman2017}. For large optical depth, most of the coherent drive is converted to an incoherent field via multiple scattering mechanism. A total conversion of field leads to $s_i=s_c$ which gives an upper bound for the modification of $\alpha=1+s_i$.
	
	For large saturations, the atom is driven in the Mollow triplet regime, and we expect this large value of $\alpha$ to be mitigated by a reduction of the reabsorption cross section \cite{Mollow1972}. In other words, the converted light will be partly off-resonant with the atomic transition.

	\section{Conclusion} \label{sec:conc}
	
	In this paper, we have thoroughly studied the total and coherent scattering rates of a MLS atoms illuminated by different configurations of the electromagnetic field that correspond to situations often encountered in experimental realization. We proposed to map these scattering properties to the one of an effective TLS model which is particularly relevant to reduce the complexity of multiple scattering simulations. We have shown that, at zero magnetic field, our effective TLS model describes exactly the scattering rates of a MLS for any saturation parameter.  In the presence of stray magnetic fields that lift the Zeeman degeneracy, the amplitude of the scattered fields is well described by the TLS model, with an exact mapping for the specific case of $\sigma$ and $\pi$ polarizations. For other polarizations, vectorial scattering (magneto-optical effects) can occur. Our scalar model cannot exactly render such rotations but proved to be robust for $\sigma$ polarized scattering. In the limit of strong saturation, incoherent scattering dominates. In dense ensembles, the dynamic of a single atom will be affected by this incoherent electromagnetic background. We have shown that, while the MLS dynamic becomes complex, an equivalent TLS model can be adapted to enlighten the general behavior and main scattering response. In particular, we noticed that even though the intrinsic reduction $\alpha$ of the saturation intensity stays close to 1, the effective reduction $\alpha_{\rm eff}$ is proportional to the incoherent background intensity thus reducing the coherent scattered field. This work gives an upper bound to situations in which a TLS model applies and can be extended to any atom having a cycling transition.
	
	
	\section*{Acknowledgements} \label{sec:acknowledgements}
	The authors thank W. Guerin, R. Bachelard and K. Vinck for helpful discussions. {R.V acknowledge PhD support from the university of Bordeaux,  J-B.G and V.M. acknowledge support from the French State, managed by the FrenchNational Research Agency (ANR) in the frame of the Investments for the future Programme IdEx Bordeaux-LAPHIA (ANR-10-IDEX-03-02). This work was also supported by the ANR contract ANR-18-CE47-0001-01.
		


		\section*{Appendix} \label{sec:appendix}
		
		In the following, we briefly give the formalism used to calculate the steady-states solutions of the density matrix that were used to evaluate the atomic scattering cross section. A detailed derivation of the formalism can be found in \cite{Steck2019book}.
		
		The lowering operator between two hyperfine states $F_{g}$ and $F_{e}$ is given by:
		\begin{equation}
				\Sigma_{q} =  \sum_{\substack{F_{g},m_{g}, \\ F_{e},m_{e}}}  \lambda(F_{g},m_{g},F_{e},m_{e},q) \ket{F_{g},m_{g}}  \bra{F_{e},m_{e}},
				\label{eq:MollowLowerOp}
		\end{equation}
		where

			\begin{align}
				\lambda(F_{g},m_{g},F_{e},m_{e},q) =& (-1)^{F_{e}+J_{g}+1+I}  \sqrt{(2F_{e}+1)(2J_{g}+1)} \nonumber \\ & \braket{F_{g},m_{g}}{F_{e},m_{e};1,q} \begin{Bmatrix} J_{e} & J_{g} & 1 \\ F_{g} & F_{e} & I \end{Bmatrix},
				\label{eq:CouplingLambda}
			\end{align}
    with the curly brackets denoting the Wigner-6j symbol.\\
	The expectation values involved in the coherent and total scattering rates of Eqs. (\ref{eq:scattering_rates_multi_levels}) using Eq. ($\ref{eq:MollowLowerOp}$) are:

				\begin{align}
					\langle  \Sigma_{q}  \rangle = &  \sum_{\substack{F_{g},m_{g}, \\ F_{e},m_{e}}}\lambda(F_{g},m_{g},F_{e},m_{e},q) {\rho}_{F_{e},m_{e},F_{g},m_{g}} \nonumber\\ 
					\sum_{q} \langle \Sigma_{q}^{\dagger} \Sigma_{q}  \rangle = & (2J_{g}+1)/(2J_{e}+1) \sum_{F_{e},m_{e}} {\rho}_{F_{e},m_{e},F_{e},m_{e}}. 
					\label{eq:lowerOp_expValue}
				\end{align}
	So the coherent scattering is given by the optical coherences and the total scattering by the population of the excited state.\\
	Projecting Eq. (\ref{eq:master_equation}) onto the general states $\bra{\alpha,m_{\alpha}}$ and $\ket{\beta,m_{\beta}}$, the full master equation for an arbitrary number of hyperfine and Zeeman states reads: 
		
		\onecolumngrid
		
		\hrulefill
		
		\begin{equation}
			\begin{split}
				\dot{{\rho}}_{\alpha,m_{\alpha},\beta,m_{\beta}} =
				& - \frac{i}{2} [
				\sum_{F_{e},m_{e}}    \Omega^{*}(\alpha,m_{\alpha} ,F_{e} ,m_{e}) \delta_{\alpha g} {\rho}_{F_{e} ,m_{e} ,\beta ,m_{\beta}}   + \sum_{F_{g},m_{g}}    \Omega(F_{g} ,m_{g} ,\alpha, m_{\alpha}) \delta_{\alpha e} {\rho}_{F_{g} ,m_{g} ,\beta ,m_{\beta}} \\
				& -   \sum_{F_{g},m_{g}}     \Omega^{*}(F_{g}, m_{g} ,\beta, m_{\beta}) \delta_{e \beta}  {\rho}_{\alpha ,m_{\alpha}, F_{g}, m_{g}}  -  \sum_{F_{e},m_{e}}     \Omega(\beta ,m_{\beta} ,F_{e} ,m_{e}) \delta_{g \beta} {\rho}_{\alpha ,m_{\alpha}, F_{e} ,m_{e}}  ] \\
				& - i \left( \delta_{\alpha g}\delta_{e \beta} - \delta_{g \beta} \delta_{\alpha e} \right)     \Delta_{F_{g}F_{e}}   {\rho}_{\alpha, m_{\alpha} ,\beta, m_{\beta}}    -   i  \Delta_{\alpha,m_{\alpha},\beta,m_{\beta}}(B_{z})   {\rho}_{\alpha, m_{\alpha}, \beta ,m_{\beta}} \\
				& - \frac{\Gamma}{2} \delta_{\alpha e} {\rho}_{\alpha,m_{\alpha},\beta m_{\beta}} - \frac{\Gamma}{2} \delta_{\beta e} {\rho}_{\alpha,m_{\alpha},\beta m_{\beta}} \\
				& +  \ \delta{\alpha g} \delta{g \beta} \sum_{q,F_{e},F_{e}'} \Gamma (-1)^{-\alpha-\beta} (2J_{e}+1)  \begin{Bmatrix} J_{g} & J_{e} & 1 \\ F_{e} & \alpha & I \end{Bmatrix} \begin{Bmatrix} J_{g} & J_{e} & 1 \\ F_{e}' & \beta & I \end{Bmatrix}  \sqrt{2\alpha+1} \sqrt{2\beta+1}   \\
				& \braket{F_{e},m_{\alpha}-q}{\alpha,m_{\alpha};1,-q}   \braket{F_{e}',m_{\beta}-q}{\beta,m_{\beta};1,-q} {\rho}_{F_{e},m_{\alpha}-q,F_{e}',m_{\beta}-q}.
			\end{split} 
			\label{eq:MEsingleModeField}
		\end{equation}
		
		\hrulefill
		\twocolumngrid
		The first 4 terms in Eq. (\ref{eq:MEsingleModeField}) are the field terms, followed by the laser detuning, the Zeeman splitting, and finally 3 decay terms proportional to $\Gamma$. ${\rho}_{i,m_{i},j,m_{j}}$ are the density matrix elements, $\delta_{i,j}$ is the Kronecker symbol, $\Delta_{F_{g},F_{e}}=\omega-\omega_{Fg,Fe}$ is the detuning of the laser compared to the atomic hyperfine transition from $F_{g}$ to $F_{e}$. The magnetic field is along $\boldsymbol\epsilon_{z}$, the Land\'{e} factors are $g_{i}$. The Zeeman shift is given by $\Delta_{\alpha,m_{\alpha},\beta,m_{\beta}}(B_{z})= \frac{\mu_{b}B_{z}}{\hbar} \left(  g_{\alpha} m_{\alpha}  -  g_{\beta} m_{\beta}  \right)$. The Rabi frequency depends on Clebsch-Gordan coefficients and the field complex amplitude:

			\begin{align} 
				\Omega(F_{e},m_{e},F_{g},m_{g})=&(-1)^{F_{e}+J_{g}+1+I} \sqrt{{(2F_{e}+1)(2J_{g}+1)}} \nonumber\\
				&\bra{F_{g},m_{g}}\ket{F_{e}, m_{e} ; 1 , m_{g}-m_{e}} \nonumber\\
				&\begin{Bmatrix} J_{g} & J_{e} & 1 \\ F_{e} & F_{g} & I \end{Bmatrix}  \Omega_{m_{g}-m_{e}}^{(J_{g},J_{e})},
				\label{eq:RabiFull}
			\end{align}
		where $\Omega_{q}^{(J_{g},J_{e})}=-{2 \bra{J_{g}}\mid{e\mathbf{r}}\mid \ket{J_{e}} E_{0q}^{(+)}}/{\hbar} $ with $\bra{J_{g}}|e\mathbf{r}| \ket{J_{e}}$ being the reduced dipole matrix element between the states $J_{g}$ and $J_{e}$ and $E_{0q}^{+}$ the positive rotating amplitude of the electric field. The coupling between two Zeeman states is expressed with the Wigner-6j symbol and Clebsch-Gordan coefficients expressed with the Wigner-3j symbols:

			\begin{align}
				\bra{F_{g},m_{g}}\ket{F_{e},1;m_{e},q} =& (-1)^{F_{e}-1+m_{g}} \sqrt{2F_{g}+1} \nonumber \\ & \begin{pmatrix} F_{e} & 1 & F_{g} \\ m_{e} & q & -m_{g} \end{pmatrix},
				\label{eq:MEsingleModeFieldDissi0}\\
				\bra{F_{e},m_{e}}\ket{F_{g},1;m_{g},-q}  =&   (-1)^{F_{g}-m_{e}-1} \sqrt{2F_{e}+1} \nonumber \\  &  \begin{pmatrix} F_{e} & 1 & F_{g} \\ m_{e} & q & -m_{g} \end{pmatrix}.
				\label{eq:MEsingleModeFieldDissi2}
			\end{align}	
		The optical coherence, under the adiabatic approximation where the coherences are always in equilibrium with respect to the population evolution ($\dot{{\rho}}_{eg}\approx0$), for a transition between $m_{e}$ and $m_{g}$ on resonance is given by (\ref{eq:optical_coherences_eg}). It depends on the population difference and on Zeeman coherences. In the case of a $pure$ polarization (aligned with only one element of the spherical basis), if the Zeeman coherences start at zero then they remain zero at steady-state, so the optical coherences are directly given by the population difference. Otherwise, Zeeman coherences can be driven and contribute to the scattering rates. 
		\begin{equation} 
			\begin{split}
				{\rho}_{e,m_{e},g,m_{g}} = & -\frac{i}{\Gamma} \sum_{m_{g}'} \Omega(m_{g}',m_{e}) {\rho}_{g,m_{g}',g,m_{g}} \\ 
				& - \frac{i}{\Gamma} \sum_{m_{e}'} \Omega(m_{g}',m_{e}') {\rho}_{e,m_{e},e,m_{e}'}.
			\end{split}
			\label{eq:optical_coherences_eg}
		\end{equation}
		To include an incoherent field background in the master equation, Zeeman coherences are neglected in Eq. (\ref{eq:optical_coherences_eg}). Writing a master equation only for incoherent field terms leads to the following rate equations with saturation $s_{i}=2\Omega_{i}^{2}/\Gamma^{2}$:		
		\begin{equation}
			\label{eq:ME_incoherent_terms}
			\begin{split}
				\Gamma \dot{{\rho}}_{\alpha,m_{\alpha},\alpha,m_{\alpha}} =
				& \delta_{\alpha e}  \sum_{i}    \Omega_{i}^{2}(m_{\alpha},i) {\rho}_{g,i,g,i}   \\
				& - \delta_{\alpha e} \left( \sum_{i}    \Omega_{i}^{2}(m_{\alpha},i) \right) {\rho}_{\alpha,m_{\alpha},\alpha,m_{\alpha}} \\
				& \delta_{\alpha g}  \sum_{i}    \Omega_{i}^{2}(i,m_{\alpha}) {\rho}_{e,i,e,i}   \\
				& - \delta_{\alpha g} \left( \sum_{i}    \Omega_{i}^{2}(i,m_{\alpha}) \right) {\rho}_{\alpha,m_{\alpha},\alpha,m_{\alpha}}. \\
			\end{split} 
			\end{equation}
		
		Finally, the full master equation including all effects is obtained by adding the terms of Eq. (\ref{eq:ME_incoherent_terms}) to Eq. (\ref{eq:MEsingleModeField}). Also, solving the full master equation at zero magnetic field, without coherent driving field ($s_c=0$) and keeping the decay terms, one obtains the total excited state population given in the main text in Eq. (\ref{eq:incoherent_tot_pop}).
		
		\vspace{1cm}
		
		\bibliography{21032020}

	\end{document}